\documentclass[conference]{IEEEtran}
   \usepackage[pdftex]{graphicx}
\usepackage[cmex10]{amsmath}
\usepackage{pgf,tikz}
\usetikzlibrary{arrows}
\usepackage{research5IEEE}

\usepackage{amssymb}
\usepackage{color}
\usepackage{bbm}

\newcommand{\len}{\textrm{len}}

\allowdisplaybreaks

\usepackage{array}

\usepackage{cite}

\newtheorem{theorem}{Theorem}

\newtheorem{definition}{Definition}
\newtheorem{lemma}{Lemma}

\renewcommand{\and}{\textnormal{and}}

\newcommand{\str}{\textrm{string}}
\newcommand{\M}{\mathsf{M}}
\newcommand{\m}{\mathsf{m}}
\begin{document}

\title{Distributed Hypothesis Testing with Variable-Length Coding}

\author{$^{\small{1}}$Sadaf Salehkalaibar and $^{\small{2}}$Mich\`{e}le Wigger \\
	\small 
	$^{1}$ECE Department, College of Engineering, University of Tehran, Tehran, Iran, s.saleh@ut.ac.ir\\
	$^{2}$LTCI, Telecom Paris, IP Paris, 75013 Paris, France,
	michele.wigger@telecom-paristech.fr
}

\maketitle

\begin{abstract} 

This paper  characterizes the optimal type-II error exponent for a  distributed hypothesis testing-against-independence problem when the \emph{expected} rate of the sensor-detector link is constrained.  Unlike for the well-known Ahlswede-Csiszar result that holds under a \emph{maximum} rate constraint and where a strong converse holds, here the optimal exponent depends on the allowed type-I error exponent.  Specifically, if the type-I error probability is limited by $\epsilon$, then the optimal type-II error  exponent under an \emph{expected} rate constraint $R$  coincides with the optimal type-II error exponent under a \emph{maximum} rate constraint of $(1-\epsilon)R$.

\end{abstract}


% \fi
\IEEEpeerreviewmaketitle

\section{Introduction}

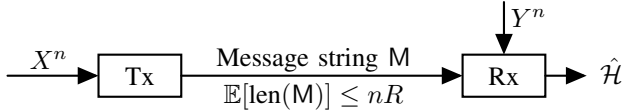
\begin{figure}[ht]
 	\vspace*{0mm}
 	\centerline{
 		\begin{tikzpicture}[line cap=round,>=triangle 45,x=1.0cm,y=1.0cm,xscale=1.1,yscale=0.7,thick]
 		%\draw[help lines]{(0,0)grid(8,3)};
 		\node at (1.5,0.8){$X^n$};	
 		\draw [->][thick](1,0.5)--(2.1,0.5);
 		%Transmitter	
 		\draw [thick] (2.1,0.1) rectangle (3.1,0.9);
 		\node at (2.6,0.5){Tx};
 		\draw [->][thick](3.1,0.5)--(6.5,0.5);
 		\node at (4.7,0.8){Message string $\mathsf{M}$};
 		\node at (4.7,0.1){ $\mathbb{E}[\len(\mathsf{M})]\leq nR$};
 		%Receiver
 		\draw [thick] (6.5,0.1) rectangle (7.5,0.9);
 		\node at (7,0.5){Rx};
 		\draw [thick][->] (7,1.9)--(7,0.9);
 		\node at (7.3,1.6){$Y^n$};
 		%Testing
 		\draw [thick][->] (7.5,0.5)--(8,0.5);
 		\node at (8.3,0.6){$\set{\hat{H}}$};
 		\end{tikzpicture}
 		\vspace*{-3mm}
 	}	
 	\caption{Variable-length hypothesis testing.\label{fig1}}
 \end{figure}

Consider the distributed hypothesis testing problem in Figure~\ref{fig1} with  a sensor  and a detector observing the source sequences  $X^n$ and $Y^n$, and where the sensor can send a bit string $\M\in\{0,1\}^*$ to the detector. The joint distribution depends on  one of two possible hypotheses,  $\mathcal{H}=H_0$ or $\mathcal{H}=H_1$, and the detector  has to decide based on $Y^n$ and $\M$ which of the two hypotheses is valid.  %The sensor can send a sequence of bits over a noise-free bit-pipe to the detector, which based on this received sequence and the observed source sequence decides on the hypothesis.
 There are two error events:  a type-I error indicates that the detector declares  $\hat{\mathcal{H}}=H_1$ when the correct hypothesis is $\mathcal{H}=H_0$, and a type-II error indicates that the detector declares $\hat{\mathcal{H}}=H_0$ when the correct hypothesis is $\mathcal{H}=H_1$. 
The goal is to maximize the exponential decay (in the blocklength $n$) of the type-II error probability under a constrained type-I error probability. The main difference of this work compared to previous works  \cite{Ahlswede, Han, Amari, Wagner, Weinberg}  is on the   constraint imposed on the communication rate. While all previous works  have constrained the \emph{maximum} number of bits that the sensor can send to the detector,  here we only constrain the \emph{expected} number of bits. Our problem is thus a relaxed version of these previous works, and can be thought of as their variable-length coding counterpart. 

In this paper, we specifically consider the distributed  \emph{testing-against-independence}   problem introduced in \cite{Ahlswede} where under the alternative hypothesis ($\mathcal{H}=H_1$) the joint distribution factorizes into the product of the marginals under  the null hypothesis ($\mathcal{H}=H_0$). The proposed setup can be considered as the variable-length extension of \cite{Ahlswede} thanks to the relaxed constraint on the expected number of communicated bits. The following strategy was proposed by Ahlswede and Csiszar and was shown to be optimal \cite{Ahlswede} under a maximum rate constraint.
The transmitter compresses its  observed source sequence  $X^n$ and describes this compressed version to the detector. If the compression fails, it sends a $0$-bit to indicate this failure. The detector  decides on $\hat{\mathcal{H}}=H_1$, whenever it receives the single $0$-bit  or   the \emph{joint  type} (the empirical symbol frequencies) of the compressed sequence and the observation $Y^n$  is not close to the one expected under $H_0$. Otherwise it decides on $\hat{\mathcal{H}}=H_1$. Notice that with the described strategy,  the type-I error probability can be made arbitrarily small as the blocklength $n$ increases. 

While optimal under a maximum rate constraint, a strategy with vanishing type-I error probability has to be wasteful under an expected rate constraint. The sensor should rather identify a subset of source  sequences $\mathcal{S}_n \subseteq \mathcal{X}^n$ of probability  close to $\epsilon$  and send a $0$-bit  whenever the observed source sequence $X^n \in \mathcal{S}_n$. In all other cases, the sensor should employ the Ahlswede-Csiszar strategy \cite{Ahlswede}  that is optimal under the maximum rate-constraint, and so should the detector. In particular, the detector should produce $\hat{\mathcal{H}}=H_1$ whenever it receives the single $0$-bit. Compared to the Ahlswede-Csiszar strategy, this new strategy achieves the same type-II error exponent; it increases the type-I   error probability by at most   $\epsilon$; and it has expected rate at most equal to $(1-\epsilon)$ times the maximum rate of the Ahlswede-Csiszar strategy. 

By means of an information-theoretic converse that uses  the \emph{$\eta$-image characterization} technique of \cite{Ahlswede}, \cite{TianChen} and the \emph{change of measure} method of \cite{Tyagi}, we show that the described strategy achieves the optimal type-II error exponent under an expected rate constraint. 
The optimal type-II error exponent under an expected rate constraint $R$ coincides with the optimal exponent under a maximum rate-constraint $(1-\epsilon)R$, when $\epsilon\in(0,1)$ denotes the allowed type-I error probability. This result  implies that under an expected rate constraint the optimal type-II error exponent depends on the allowed type-I error probability and  a strong converse like under a maximum rate-constraint does not hold.

%The required rate of the noiseless link is $1-\epsilon$ times that of the fixed-length setup where $\epsilon$ is the threshold on the type-I error probability. To highlight the difference of the two setups, the idea of the variable-length coding scheme is given as follows. The transmitter performs a typicality test on the observed sequence. If it is not typical according to $P_X$ or it is typical but the probability of its occurance is smaller than $\epsilon$,  a \emph{zero-bit} is sent to the receiver, otherwise, the compressed codeword is communicated. Moreover, the variable-length coding allows the system  require a smaller rate thanks to the transmission of a zero-bit in the case of \emph{atypicality}. 

%We conclude this section with some remarks on the notation. 

\subsection{Notation}

We mostly follow the notation in \cite{ElGamal}. For a given pmf $P_X$  the set of sequences whose type (symbol frequencies) is described by $P_X$ \cite{cover} is denoted by $\mathcal{T}^n(P_X)$. For a given $P_X$ and small number $\mu>0$, the set of all sequences in $\mathcal{X}^n$ whose type has $\ell_1$-distance from $P_X$ at most equal to $\mu$ is called the $\mu$-typical set around $P_X$ and is denoted  $\mathcal{T}_{\mu}^n(P_X)$.

For any positive integer number $m\geq 1$, we use $\str(m)$ to denote the bit-string of length $\lceil \log_2(m)\rceil$ representing $m$. We further use sans serif font to denote bit-strings of arbitrary lengths: for example $\mathsf{m}$ for a deterministic bit-string and $\mathsf{M}$ for a random bit-string. The function $\len(\mathsf{m})$ returns the length of a given bit-string $\mathsf{m}\in\{0,1\}^*$.

%\section{Distributed Hypothesis Testing Over a Positive-Rate Noiseless Link}\label{sec:noiseless}
\section{System Model}

Consider the distributed hypothesis testing problem with a transmitter and a receiver in Fig.~\ref{fig1}. The transmitter observes the source sequence $X^n$ and the receiver observes the source sequence $Y^n$. Under the null hypothesis 
\begin{align}
	\mathcal{H}=H_0\colon \quad (X^n,Y^n)\sim \text{i.i.d.}\; P_{XY},
\end{align}
for a given pmf $P_{XY}$, whereas under the alternative hypothesis 
\begin{align}
	\mathcal{H}=H_1\colon \quad (X^n,Y^n)\sim \text{i.i.d.}\; P_{X}\cdot P_Y.
\end{align}

There is a noise-free bit pipe from the transmitter to the receiver. Upon observing $X^n$, the transmitter computes the message $\M=\phi^{(n)}(X^n)$ using a possibly stochastic encoding function \begin{align}\phi^{(n)}: \set{X}^n\to \{0,1 \}^*,\end{align} such that\footnote{The expectation in \eqref{L-def} is with respect to the law of $X^n$ which equals $P_X^n$ under both hypotheses.} \begin{align}\mathbb{E}\left[\len(\M)\right]\leq nR.\label{L-def}\end{align}  It then sends a bitstring $\M$ over the bit pipe to the receiver.  

The goal of the communication is that the receiver can determine the hypothesis $\mathcal{H}$ based on its observation $Y^n$ and its received message. Specifically, the receiver  produces the guess \begin{equation}\hat{\mathcal{H}}=g^{(n)}(Y^n,\M)
\end{equation}
using a decoding function $g^{(n)}:\mathcal{Y}^n\times \{0,1\}^*\to \{H_0,H_1\}$.  This induces a partition of the sample space $\set{X}^n\times \set{Y}^n$ into an acceptance  region $\mathcal{A}_n$  for hypothesis  $H_0$,
\begin{align}
	&\mathcal{A}_n \triangleq  \big\{(x^n,y^n)\colon  g^{(n)}(y^n,\phi^{(n)}(x^n))=H_0 \big\},
\end{align}
and a rejection region for $H_0$: \begin{align}\mathcal{A}^c_n\triangleq (\mathcal{X}^n\times \mathcal{Y}^n)\backslash \mathcal{A}_n.
\end{align}

\begin{definition}\label{def} For any $\epsilon \in [0,1)$ and for a given rate $R\in \Reals_+$,  a type-II exponent $\theta\in \Reals_+$ is $(\epsilon,R)$-achievable 
	if there exists a sequence of functions $(\phi^{(n)},g^{(n)})$, such that the corresponding sequences of type-I  error probability 
	\begin{align}\alpha_{n}\eqdef P_{XY}^n(\set{A}_n^c)\end{align} and type-II error probability \begin{align}\beta_{n}\eqdef P_{X}^nP_{Y}^n(\set{A}_n),\end{align}
	respectively, satisfy 
	\begin{align}
		\alpha_{n}&\leq \epsilon,\label{type-I-def}
	\end{align}
	and
	\begin{align}
	\liminf_{n\to\infty}\;\frac{1}{n}\log\frac{1}{\beta_{n}}\geq \theta.\label{type-II-cons}
	\end{align}
	The optimal exponent $\theta_\epsilon^*(R)$ is the supremum of all $(\epsilon,R)$-achievable type-II exponents $\theta\in \Reals_+$.
\end{definition}

\section{Optimal Error Exponent}

%The following theorem establishes the optimal error exponent $\theta_{\epsilon}^*(R)$.

\begin{theorem}\label{thm1} The optimal exponent is given by 
	\begin{align}
	\theta^*_{\epsilon}(R) = \max_{\substack{P_{U|X}\colon \\R\geq (1-\epsilon)I(U;X)}}I(U;Y).
	\end{align}
	where the mutual informations are  evaluated with respect to the joint pmf 
	\begin{align}\label{eq:law}
	P_{UXY}\triangleq P_{U|X}\cdot P_{XY}.
	\end{align}
	\end{theorem}
\begin{IEEEproof} Here we only prove  achievability. The converse is proved in Section~\ref{sec:converse}.   
	
	\textbf{Achievability}: Fix a large blocklength $n$, a small number $\mu\in(0, \epsilon)$, and a conditional pmf $P_{U|X}$ such that:
	\begin{align}\label{eq:rateR}
	R= (1-\epsilon+\mu)I(U;X)+\mu,
	\end{align}
	where mutual is evaluated according to the pmf in \eqref{eq:law}. Randomly generate an $n$-length  codebook $\mathcal{C}_U$ of rate $R$  
	 by picking all entries i.i.d. according to the marginal pmf $P_U$. The realization of the codebook 
	 		\begin{align}
	 		\mathcal{C}_U\triangleq \left\{ u^n(m)\colon m\in\left\{1,\ldots,\lfloor 2^{nR}\rfloor\right\}  \right\}
	 		\end{align} is revealed to all terminals.

	Finally, choose a subset $\mathcal{S}_n\subseteq \mathcal{T}_{\mu}^{(n)}(P_X)$ such that \begin{equation}
	\Pr\left[X^n\in \mathcal{S}_n\right]=\epsilon-\mu.
	\end{equation}

%	 where $\xi_n$ is any function that  goes to zero as $n\to\infty$. 
	
	\underline{\textit{Transmitter}}: Assume it observes $X^n=x^n$. If 
	\begin{equation}
	x^n \notin \mathcal{S}_n,
	\end{equation}
	 it looks for an index $m$ such that 
\begin{equation}(u^n(m),x^n)\in\mathcal{T}_{\mu}^n(P_{UX}). 
\end{equation}If  successful, it picks  one of these indices  uniformly at random and sends  the binary representation of length $\lceil \log_2(m^\star)\rceil$ of the chosen index  $m^\star$ over the noiseless link: %. So, if the chosen index is $m^*$, it sends the corresponding bit-string
\begin{equation}\M = \str(m^\star).
\end{equation}
Otherwise it sends the single bit $\M=[0]$. 
	
	\underline{\textit{Receiver}}: If it  receives  the single bit $\M=[0]$, it declares $\hat{\mathcal{H}}=H_1$. Otherwise, it converts the received bit string $\M$ into an index $m$ and  checks whether $(u^n(m),y^n)\in\mathcal{T}_{\mu}^n(P_{UY})$. If successful, it declares $\hat{\mathcal{H}}=H_0$, and otherwise it declares $\hat{\mathcal{H}}=H_1$.
	
	\underline{\textit{Analysis}}:
	Since a single bit is sent when $x^n \in \mathcal{S}_n$ and since never more than $n(I(U;Y)+\mu)$ bits are sent, the expected message length can be bounded  as:
	\begin{IEEEeqnarray}{rCl}
		\mathbb{E}\left[\len(\M)\right] &= & \Pr[X^n\in \mathcal{S}_n]\cdot \mathbb{E}\left[\len(\M)|X^n\in \mathcal{S}_n\right]\nonumber\\&&\hspace{0.5cm}+\Pr[X^n\notin \mathcal{S}_n]\cdot \mathbb{E}\left[\len(\M)|X^n\notin \mathcal{S}_n\right]\\\nonumber\\
		&\leq& (\epsilon-\mu)\cdot 1 + (1-\epsilon+\mu)\cdot n(I(U;X)+\mu),\nonumber\\
	\end{IEEEeqnarray}
which for sufficiently large $n$ is further bounded as (see  \eqref{eq:rateR}):
\begin{equation}\label{eq:LR}
		\mathbb{E}\left[\len(\M)\right] < nR.
\end{equation}

To bound the type-I and type-II error probabilities, we notice that when $x^n \notin \mathcal{S}_n$, the scheme  coincides with the one proposed by Ahlswede and Csisz\`ar in \cite{Ahlswede}. When $x^n \in \mathcal{S}_n$, the transmitter sends the single bit $\M=[0]$ and the receiver declares $H_1$. The type-II error probability of our scheme is thus no larger than the type-II error probability of the scheme in \cite{Ahlswede}, and the type-I error probability is at most 
$\Pr[X^n\in \mathcal{S}_n]=\epsilon-\mu$ larger than in \cite{Ahlswede}. 
Since the type-I error probability in \cite{Ahlswede} tends to 0 as $n \to \infty$,
 the type-I error probability here is bounded by
 $\epsilon$, for sufficiently large values of $n$ and all choices of $\mu \in(0,\epsilon)$.
 Combining  the result in \cite{Ahlswede}, with  \eqref{eq:LR}, and  letting $\mu \to 0$ thus establishes the achievability part of the proof.
For the converse proof see the following Section~\ref{sec:converse}.
	\end{IEEEproof}

\section{Proof of Converse to Theorem~\ref{thm1}}\label{sec:converse}
Fix an achievable exponent $\theta<\theta^*_{\epsilon}(R)$ and a sequence of encoding and decision functions so that \eqref{type-I-def} and \eqref{type-II-cons} are satisfied. Fix also an integer $n$ and a small number $\eta\geq 0$ and define the set
\begin{IEEEeqnarray}{rCl}\label{eq:Bn}
	\mathcal{B}_n(\eta) \triangleq \left\{ x^n\colon \Pr\Big[\hat{\mathcal{H}}=H_0\Big |X^n=x^n,\mathcal{H}=H_0\Big]\geq \eta \right\}.\nonumber\\
\end{IEEEeqnarray}
Notice that by the constraint on the type-I error probability, \eqref{type-I-def},
\begin{IEEEeqnarray}{rCl}
	1-\epsilon &\leq & 
	\sum_{x^n\in\mathcal{B}_n(\eta)}\Pr\Big[\hat{\mathcal{H}}=H_0\Big|X^n=x^n,\mathcal{H}=H_0\Big] P_X^n(x^n)\nonumber\\&&\hspace{0cm}+\sum_{x^n\notin\mathcal{B}_n(\eta)}\Pr\Big[\hat{\mathcal{H}}=H_0\Big |X^n=x^n,\mathcal{H}=H_0\Big]  P_X^n(x^n) \nonumber\\\\
	&\leq &P_X^n(\mathcal{B}_n(\eta))+\eta(1-P_X^n(\mathcal{B}_n(\eta))).
\end{IEEEeqnarray}	
Thus, 
\begin{IEEEeqnarray}{rCl}
	P_X^n(\mathcal{B}_n(\eta))\geq \frac{1-\epsilon-\eta}{1-\eta}.\label{step1} 
\end{IEEEeqnarray}

Define now 
\begin{align}
\mu_n\triangleq n^{-\frac{1}{3}}
\end{align} and 
\begin{align}\mathcal{D}_n(\eta)\triangleq \mathcal{T}_{\mu_n}^{n}(P_X)\cap \mathcal{B}_n(\eta).\end{align} 
By \cite[Lemma 2.12]{Csiszarbook}:
\begin{IEEEeqnarray}{rCl}
	P_X^n(\mathcal{T}_{\mu_n}^{n}(P_X)) \geq 1-\frac{|\mathcal{X}|}{2\mu_nn},\label{step2}
\end{IEEEeqnarray}which combined with 
\eqref{step1} and  the general identity $\Pr(A\cap B)\geq \Pr(A)+\Pr(B)-1$ yields:
\begin{IEEEeqnarray}{rCl}\label{eq:Deltan}
	P_X^n(\mathcal{D}_n(\eta)) \geq \frac{1-\epsilon-\eta}{1-\eta}-\frac{|\mathcal{X}|}{2\mu_nn}\triangleq \Delta_n.
\end{IEEEeqnarray}

Define  the random variables $(\tilde{\M},\tilde{X}^n,\tilde{Y}^n)$ as the restriction of the triple $(\M, X^n, Y^n)$ to  $X^n \in \mathcal{D}_n(\eta)$. The probability distribution of the restricted triple  is then given by:
\begin{IEEEeqnarray}{rCl}
	&&P_{\tilde{\M}\tilde{X}^n\tilde{Y}^n}(\mathsf{m},x^n,y^n)\triangleq\nonumber\\&&\hspace{0.3cm} P_{XY}^n(x^n,y^n)\cdot \frac{\mathbbm{1}\left\{ x^n\in\mathcal{D}_n(\eta) \right\}}{P_X^n(\mathcal{D}_n(\eta))}\cdot \mathbbm{1}\left\{ \phi^{(n)}(x^n)=\mathsf{m} \right\}.\IEEEeqnarraynumspace \label{tilde-def}
\end{IEEEeqnarray}
This implies in particular:
\begin{IEEEeqnarray}{rCl}
	P_{\tilde{X}^n}(x^n)&\leq& P_X^n(x^n)\cdot 
	\Delta_n^{-1},\label{step11}\\
	P_{\tilde{Y}^n}(y^n)&\leq& P_Y^n(y^n)\cdot 
	\Delta_n^{-1},\label{mar1}\\
	P_{\tilde{\M}}(\mathsf{m}) &\leq & P_{\M}(\mathsf{m})\cdot \Delta_n^{-1},
	\label{mar2}
\end{IEEEeqnarray}
and
\begin{IEEEeqnarray}{rCl}
	D\left(P_{\tilde{X}^n}\|P_X^n\right)\leq \log
	\Delta_n^{-1}. 
\end{IEEEeqnarray}

\noindent\underline{\textit{Single-letter characterization of the rate constraint}}: Define the random variables ${L}\triangleq \len({\M})$ and  $\tilde{L}\triangleq \len(\tilde{\M})$, and notice that by the rate constraint \eqref{L-def}:
\begin{IEEEeqnarray}{rCl}
	nR &\geq & \mathbb{E}\left[L\right]\\
	&=& \mathbb{E}\left[L|X^n\in\mathcal{D}_n(\eta)\right]\cdot P_X^n(\mathcal{D}_n(\eta))\nonumber\\&&\hspace{0.5cm}+\mathbb{E}\left[L|X^n\notin\mathcal{D}_n(\eta)\right]\cdot (1-P_X^n(\mathcal{D}_n(\eta)))\\
	&\geq & \mathbb{E}\left[L|X^n\in\mathcal{D}_n(\eta)\right]\cdot P_X^n(\mathcal{D}_n(\eta))\\
	&= & \mathbb{E}\left[\tilde{L}\right]\cdot P_X^n(\mathcal{D}_n(\eta))\label{step60}\\
	&\geq & \mathbb{E}\left[\tilde{L}\right]\cdot \Delta_n,\label{step3}
\end{IEEEeqnarray}
where \eqref{step60} holds because $\tilde{\M}$ is obtained by restricting $\M$ to the event $X^n \in \mathcal{D}_n(\eta)$ and $\tilde{L}$ denotes the length of $\tilde{\M}$; and step~\eqref{step3} holds by the definition of $\Delta_n$ in \eqref{eq:Deltan}.

Now, since $\tilde{L}$ is  function of $\tilde{\M}$, we  have:
\begin{IEEEeqnarray}{rCl}
	H(\tilde{\M}) &=& H(\tilde{\M},\tilde{L})\\
	&=& H(\tilde{\M}|\tilde{L})+H(\tilde{L})\\
	&=& \sum_{\ell} \Pr(\tilde{L}=\ell)H(\tilde{\M}|\tilde{L}=\ell)+H(\tilde{L})\\
	&\leq & \sum_{\ell} \Pr(\tilde{L}=\ell)\ell+H(\tilde{L}) \label{eq:step_before4}\\
	&=& \mathbb{E}[\tilde{L}]+H(\tilde{L})\\
	&\leq & \frac{nR}{\Delta_n}+H(\tilde{L})\label{step4}\\
	&\leq &  \frac{nR}{\Delta_n}+\frac{nR}{\Delta_n}h_{\text{b}}\left(\frac{\Delta_n}{nR}\right)\label{step5}\\
	&=& \frac{nR}{\Delta_n}\left(1+h_{\text{b}}\left(\frac{\Delta_n}{nR}\right)\right).
	\label{step12}
\end{IEEEeqnarray}
Here, \eqref{eq:step_before4} holds because when  $\M$ consists of $\ell$ bits ($L=\ell$), then its entropy cannot exceed $\ell$;   \eqref{step4} follows from \eqref{step3}; and \eqref{step5} holds because when $\mathbb{E}[\tilde{L}]\leq \frac{nR}{\Delta_n}$, then the entropy of $\tilde{L}$ can be at most that of a Geometric distribution with mean $\frac{nR}{\Delta_n}$, which is $\frac{nR}{\Delta_n}\cdot h_{\text{b}}\left(\frac{\Delta_n}{nR}\right)$.

On the other hand, we  can  lower bound $H(\tilde{\M})$ in the following way: 
\begin{IEEEeqnarray}{rCl}
	&&\hspace{-0.5cm}H(\tilde{\M}) \nonumber\\&\geq & I(\tilde{\M};\tilde{X}^n)\label{eq1}\\
	&=& H(\tilde{X}^n)-H(\tilde{X}^n|\tilde{\M})\\
	&=& -\sum_{x^n} P_{\tilde{X}^n}(x^n)\log P_{\tilde{X}^n}(x^n)-H(\tilde{X}^n|\tilde{\M})\\
	&\geq& -\sum_{x^n} P_{\tilde{X}^n}(x^n)\log P_{X^n}(x^n)+\log \Delta_n-H(\tilde{X}^n|\tilde{\M})\label{step6}\\
	&=& -\sum_{x^n} P_{\tilde{X}^n}(x^n)\sum_{t=1}^n\log P_{X}(x_t)+\log \Delta_n-H(\tilde{X}^n|\tilde{\M})\nonumber\\\label{step7}\\
	%	&=& -\sum_{t=1}^n\sum_{x^n} P_{\tilde{X}^n}(x^n)\log P_{X}(x_t)+\log \Delta_n-H(\tilde{X}^n|\tilde{\M})\\
	&=& -\sum_{t=1}^n\sum_{x_t} P_{\tilde{X}_t}(x_t)\log P_{X}(x_t)+\log \Delta_n\nonumber\\&&\hspace{1cm}-H(\tilde{X}^n|\tilde{\M})\\
	&= &\sum_{t=1}^nH(\tilde{X}_t)+\sum_{t=1}^n D(P_{\tilde{X}_t}\|P_{X})+\log \Delta_n\nonumber\\&&\hspace{1cm}-H(\tilde{X}^n|\tilde{\M})\label{step8}\\
	&=&\sum_{t=1}^n\left[H(\tilde{X}_t)-H(\tilde{X}_t|\tilde{M},\tilde{X}^{t-1})\right]\nonumber\\&&\hspace{1cm}+\sum_{t=1}^n D(P_{\tilde{X}_t}\|P_{X})+\log \Delta_n\\
	&=&\sum_{t=1}^nI(\tilde{U}_t;\tilde{X}_t)+\sum_{t=1}^n D(P_{\tilde{X}_t}\|P_{X})+\log \Delta_n\label{step9}\\
	&=&nI(\tilde{U}_T;\tilde{X}_T|T)\nonumber\\&&\hspace{1cm}+\sum_{t=1}^n\;\sum_{x\in\mathcal{X}} P_{\tilde{X}_T|T=t}(x)\log \frac{P_{\tilde{X}_T|T=t}(x)}{P_{X}(x)}+\log\Delta_n\nonumber\\\\
	&=& nI(\tilde{U}_T;\tilde{X}_T|T)\nonumber\\&&\hspace{1cm}+\sum_{t=1}^n\;\sum_{x\in\mathcal{X}} P_{\tilde{X}_T|T=t}(x)\log \frac{P_{\tilde{X}_T|T=t}(x)}{P_{\tilde{X}_T}(x)}\nonumber\\&&\hspace{1cm}+\sum_{t=1}^n\;\sum_{x\in\mathcal{X}} P_{\tilde{X}_T|T=t}(x)\log \frac{P_{\tilde{X}_T}(x)}{P_{X_t}(x)}+\log \Delta_n\nonumber\\\\
	&=& nI(\tilde{U}_T;\tilde{X}_T|T)+nI(\tilde{X}_T;T)+nD(P_{\tilde{X}_T}\|P_{X_T})+\log \Delta_n\nonumber\\\label{step10}\\
	&\geq & nI(\tilde{U}_T,T;\tilde{X}_T)+\log \Delta_n\\
	&=&nI(U;\tilde{X})+\log \Delta_n,\label{step20}
\end{IEEEeqnarray}
where 
\begin{itemize}
	\item \eqref{step6} holds by \eqref{step11};
	\item \eqref{step7} holds because $X^n$ is i.i.d. under $P_X^n$;
	\item \eqref{step9} holds by defining $\tilde{U}_t\triangleq (\tilde{\M},\tilde{X}^{t-1})$;
	\item \eqref{step10} holds because $T$ is uniformly chosen over $\{1,\ldots,n\}$;
	\item \eqref{step20} follows by defining $U\triangleq (\tilde{U}_T,T)$ and $\tilde{X}\triangleq \tilde{X}_T$.
\end{itemize}
Combining \eqref{step12} and \eqref{step20}, we obtain:
\begin{IEEEeqnarray}{rCl}\label{eq:la}
	R\geq \frac{  I(U; \tilde{X})+ \frac{1}{n} \log \Delta_n}{ 1+h_{\text{b}}\left(\frac{\Delta_n}{nR} \right)} \cdot \Delta_n.
\end{IEEEeqnarray}
%Letting now $n\to \infty$, the term $\Delta_n$ tends to $\frac{1-\epsilon-\eta}{1-\eta}$, and  the right-hand side of \eqref{eq:la}  to $I(U; \tilde{X})\cdot \frac{1-\epsilon-\eta}{1-\eta}$.
\bigskip 

\noindent\underline{\textit{Upper bounding the error exponent}}:
For each string $\m \in \{0,1\}^*$, define
 the following sets:
\begin{IEEEeqnarray}{rCl}
	\mathcal{F}_\m &\triangleq & \left\{ x^n\in\mathcal{X}^n\colon \phi^{(n)}(x^n)=\m \right\}\bigcap \mathcal{D}_n(\eta),\\	
	\mathcal{G}_\m &\triangleq & \left\{ y^n\in\mathcal{Y}^n\colon g^{(n)}(y^n,\m)=H_0 \right\}.
\end{IEEEeqnarray}
Using \eqref{mar2}, the type-II error probability can then be lower bounded as:
\begin{IEEEeqnarray}{rCl}
	\beta_n 
	&= & \sum_{\m}  P_{\M}(\m) \cdot P_Y^n(\mathcal{G}_\m)
%	& = &  \sum_{x^n,\m}  P_{X^n}(x^n)\cdot \mathbbm{1}\big\{ \phi^{(n)}(x^n)=\m\big\} \cdot P_Y^n(\mathcal{D}_\m) \\
%		& \geq  & P_X^n(\mathcal{D}_n(\eta))\cdot \sum_{x^n \in \psi_{n}(\eta)}\; \sum_{\m} P_{\tilde{X}^n}(x^n) \cdot \mathbbm{1}\big\{ \phi^{(n)}(x^n)=\m\big\} \cdot P_Y^n(\mathcal{D}_\m)\label{step31}  \\
	\geq \Delta_n \cdot\sum_\m P_{\tilde{\M}}(\m)\cdot   P_Y^n(\mathcal{G}_\m).\nonumber\\\label{step33a}
\end{IEEEeqnarray}
%where \eqref{step33a} holds by restricting the sum only over elements $x^n \in \mathcal{D}_n(\eta)$ and by the definition of the pair  $(\tilde{M},\tilde{X}^n)$
%in  \eqref{tilde-def}.

In order to find a lower bound to the right hand-side of \eqref{step33a}, we need the following definition and lemma. A set $\set{B}\subseteq \mathcal{Y}^n$ is an $\eta$-image of the set $\set{A}\subseteq \mathcal{X}^n$ if 
	\begin{IEEEeqnarray}{rCl}
		P_{Y|X}^n(\set{B}|x^n)\geq \eta,\qquad \forall x^n\in \mathcal{A}. 
	\end{IEEEeqnarray}
	The following lemma is a simple restatement of the lemma proved in \cite{TianChen}.
	\begin{lemma}[Lemma~3 in\cite{TianChen}]\label{lemma} Consider a set $\set{A}\subseteq \mathcal{X}^n$,  a number $\eta \in (0,1)$, and an $\eta$-image $\set{B}$ of $\set{A}$ with respect to the channel $P_{Y|X}$. Then, for any number $\delta'>0$ and  any output distribution  $P_{Y_{\mathcal{A}}^n}$ induced over the channel $P_{Y|X}^n$  by an arbitrary input distribution $P_\set{A}$ on $\set{A}$, i.e.,
		\begin{IEEEeqnarray}{rCl}
			P_{Y_{\set{A}}^n}(y^n)\triangleq \sum_{x^n\in \set{A}} P_{\set{A}}(x^n)P_{Y|X}^n(y^n|x^n),
		\end{IEEEeqnarray}
for all	 sufficiently large blocklengths $n$:
		\begin{IEEEeqnarray}{rCl}
P_{Y^n}(B)\geq 2^{-D\left(P_{Y_{\set{A}}^n}\big\|P_Y^n\right)-n\delta'}.
		\end{IEEEeqnarray}
	\end{lemma}
	
	To apply this lemma, we notice that the set $\mathcal{G}_m$ is an $\eta$-image of the set $\mathcal{F}_m$. In fact,  by \eqref{eq:Bn}, under $\mathcal{H}=H_0$,  whenever  $X^n \in \mathcal{D}_n(\eta)$  the receiver guesses $\hat{\mathcal{H}}=H_0$ with probability at least $\eta$.  Since  $X^n \in \mathcal{F}_{\m}$ implies   $X^n\in\mathcal{D}_n(\eta)$ and  $\M=\m$, the probability that $Y^n\in \mathcal{G}_{\m}$ needs to be at least $\eta$.
		
		We can use this observation and Lemma~\ref{lemma} to further
 lower bound the sum in~\eqref{step33a} for any $\delta'>0$ and any sufficiently large $n$:
	\begin{IEEEeqnarray}{rCl}
\lefteqn{\sum_\m P_{\tilde{\M}}(\m)\cdot  P_Y^n(\mathcal{G}_\m)} \qquad \nonumber \\
	%	&\geq& P_X^n(\mathcal{D}_n(\eta))\cdot\sum_\m P_{\tilde{\M}}(\m)\label{step51}\\
		&\geq &2^{-n\delta'} \sum_\m P_{\tilde{\M}}(\m) 2^{-D\left(P_{{\tilde{Y}^n|\tilde{\M}=\m}} \big \|P_Y^n\right)}\label{step34}\\
		&\geq& 2^{-n\delta'} 2^{-\sum_\m P_{\tilde{\M}}(\m) D\left(P_{{\tilde{Y}^n|\tilde{\M}=\m}} \big\|P_Y^n\right)}\label{step35}
	\end{IEEEeqnarray}
	where
	\begin{itemize}
	%	\item \eqref{step51} ;
		\item \eqref{step34} holds by Lemma~\ref{lemma} for the choice $A=\mathcal{F}_\m$, because $ \mathcal{G}_\m$ is an $\eta$-image of the set $\mathcal{F}_m$ and because according to \eqref{tilde-def}, $P_{\tilde{Y}^n|\tilde{\M}}(\cdot|\m)$ is the  output distribution induced by channel $P_{Y|X}^n$ for  input distribution $P_{\tilde{X}^n|\tilde{\M}}(\cdot |\m)$ over the set $\mathcal{F}_{\m}$;
		\item  \eqref{step35} holds by the convexity of the function $t\mapsto 2^t$.
	\end{itemize}
	We define $\delta''\triangleq \delta'-\frac{1}{n}\log\Delta_n$ and combine \eqref{step33a} with \eqref{step35} to  obtain: 
	\begin{IEEEeqnarray}{rCl}
		\lefteqn{-\frac{1}{n}\log \beta_n }\quad \nonumber\\&\leq & %\frac{1}{n}\sum_\m P_{\tilde{\M}}(\m) D\left(P_{{\tilde{Y}^n|\tilde{\M}}}\big\|P_Y^n\big|\tilde{\M}=\m\right)+\delta''\\
	%	&=&
	 \frac{1}{n}\sum_\m\sum_{y^n\in\mathcal{Y}^n}P_{\tilde{\M}\tilde{Y}^n}(\m,y^n)\log \frac{P_{\tilde{Y}^n|\tilde{\M}}(y^n|\m)}{P_Y^n(y^n)}+\delta''\IEEEeqnarraynumspace\\
		&=&\frac{1}{n}D(P_{\tilde{\M}\tilde{Y}^n}\|P_{\tilde{\M}} P_Y^n)+\delta''\IEEEeqnarraynumspace\label{step37}\\
%	\end{IEEEeqnarray}}
%	
%continue to upper bound the KL-divergence term $D\left(P_{\tilde{\M}\tilde{Y}^n}\|P_{\M}P_{Y}^n\right)$:
%	\begin{IEEEeqnarray}{rCl}
%		D\left(P_{\tilde{\M}\tilde{Y}^n}\|P_{\M}P_{Y}^n\right)
 &=&  \frac{1}{n} D\left(P_{\tilde{\M}\tilde{Y}^n}\|P_{\tilde{\M}}P_{\tilde{Y}^n}\right)+ \frac{1}{n}{E}_{P_{\tilde{Y}^n}}\left[\log \frac{P_{\tilde{Y}^n}}{P_{Y}^n}\right]+\delta''\\
		&\leq &	 \frac{1}{n} D\left(P_{\tilde{\M}\tilde{Y}^n}\|P_{\tilde{\M}}P_{\tilde{Y}^n}\right)+\	 \frac{1}{n}\log \Delta_n^{-1}+\delta''\label{step21}\\
		&= & 	 \frac{1}{n}I(\tilde{\M};\tilde{Y}^n)+	 \frac{1}{n}\log\Delta_n^{-1}+\delta''\\
		&=&	 \frac{1}{n} \sum_{t=1}^nI(\tilde{\M};\tilde{Y}_t|\tilde{Y}^{t-1})+	 \frac{1}{n}\log\Delta_n^{-1}+\delta''\\
		&\leq &	 \frac{1}{n}\sum_{t=1}^nI(\tilde{\M},\tilde{Y}^{t-1};\tilde{Y}_t)+	 \frac{1}{n}\log\Delta_n^{-1}+\delta''\\
		&\leq &	 \frac{1}{n}\sum_{t=1}^nI(\tilde{\M},\tilde{X}^{t-1};\tilde{Y}_t)+	 \frac{1}{n}\log \Delta_n^{-1}+\delta''\label{step22}\\
		&=&	 \frac{1}{n} \sum_{t=1}^nI(\tilde{U}_t;\tilde{Y}_t)+	 \frac{1}{n}\log \Delta_n^{-1}+\delta''\\
		&=& I(\tilde{U}_T;\tilde{Y}_T|T)+	 \frac{1}{n}\log \Delta_n^{-1}+\delta''\\
		&\leq & I(\tilde{U}_T,T;\tilde{Y}_T)+	 \frac{1}{n}\log \Delta_n^{-1}+\delta''\\
		&=& I(U;\tilde{Y})+	 \frac{1}{n}\log\Delta_n^{-1}+\delta'',\IEEEeqnarraynumspace\label{step23}
	\end{IEEEeqnarray}
	where
	\begin{itemize}
			\item \eqref{step21} holds by \eqref{mar1} and \eqref{mar2};
			\item \eqref{step22} holds by the Markov chain $\tilde{Y}^{t-1}\to (\tilde{\M},\tilde{X}^{t-1})\to \tilde{Y}_t$;
			\item \eqref{step23} follows by defining $\tilde{Y}\triangleq \tilde{Y}_T$.
		\end{itemize}
		Thus, from \eqref{step23}, we have:
		\begin{IEEEeqnarray}{rCl}
			-\frac{1}{n}\log\beta_n \leq I(U;\tilde{Y})+\frac{1}{n}\log \Delta_n^{-1}+\delta''.
		\end{IEEEeqnarray}
		
Notice that according to \eqref{tilde-def}, the distribution $P_{\tilde{X}^n}$ is a restriction to the set $\mathcal{D}_n(\eta)$ which is a subset of the typical set, thus we have $|{P}_{\tilde X}-P_X|\leq \mu_n$. Also, from $P_{\tilde{Y}|\tilde{X}}=P_{Y|X}$, and by the uniform continuity of the involved information quantities, we get as $n\to\infty$ and $\eta\to 0$:
		\begin{IEEEeqnarray}{rCl}
			R&\geq& (1-\epsilon)I(U;X),\\
			\theta &\leq& I(U;Y).
		\end{IEEEeqnarray}
		This concludes the proof of the converse.

\section{Conclusion}

We  established  the optimal type-II error exponent of a distributed testing-against-independence problem   under a constraint on the probability of type-I error and on the expected communication rate. This result can be seen as a variable-length coding version of the well-known result by Ahlswede and Csiszar \cite{Ahlswede} which holds under a maximum rate-constraint. Interestingly, the optimal type-II error  exponent under an expected rate  constraint $R$ coincides with  the optimal type-II error exponent under a maximum rate constraint $(1-\epsilon)R$ when the type-I error probability is constrained to be at most $\epsilon\in(0,1)$. Thus, unlike in the scenario with a maximum rate constraint, here the  strong converse fails because   the optimal type-II error exponent depends on the allowed type-I error probability  $\epsilon$.

\section*{Acknowledgements}
M. Wigger and S. Salehkalaibar acknowledge  funding support from the ERC under grant  agreement 715111.
\bibliographystyle{IEEEtran}
\bibliography{references}

\appendices

\end{document}